\documentclass[aps,prd,nobibnotes,twocolumn,superscriptaddress,bibliography]{revtex4-1}

\usepackage{amsfonts}
\usepackage{mathrsfs}
\usepackage{amsmath}
\usepackage{color}
\usepackage{graphicx}
\usepackage{bm}
\usepackage{amssymb}
\usepackage{xspace}
\usepackage{epstopdf}
\usepackage{dcolumn}
\usepackage{longtable}
\usepackage{multirow}
\usepackage{float}
\usepackage{comment}
\usepackage{lineno}

\usepackage[colorlinks=true, letterpaper=true, pdfstartview=FitV,  linkcolor=blue, citecolor=blue, urlcolor=blue]{hyperref}

\makeatother

\begin{document}

\title{Triply-degenerate point in three-dimensional spinless systems}

\author{Xiaolong Feng}
\email{xiaolong\_feng@mymail.sutd.edu.sg}
\address{Research Laboratory for Quantum Materials, Singapore University of Technology and Design, Singapore 487372, Singapore}

\author{Weikang Wu}
\address{Research Laboratory for Quantum Materials, Singapore University of Technology and Design, Singapore 487372, Singapore}
\address{Division of Physics and Applied Physics, School of Physical and Mathematical Sciences, Nanyang Technological University, Singapore, 637371, Singapore}

\author{Zhi-Ming Yu}
\email{zhiming\_yu@bit.edu.cn}
\affiliation{Centre for Quantum Physics, Key Laboratory of Advanced Optoelectronic Quantum Architecture and Measurement (MOE), School of Physics, Beijing Institute of Technology, Beijing, 100081, China}
\affiliation{Beijing Key Lab of Nanophotonics \& Ultrafine Optoelectronic Systems, School of Physics, Beijing Institute of Technology, Beijing, 100081, China}
\author{Shengyuan A. Yang}
\address{Research Laboratory for Quantum Materials, Singapore University of Technology and Design, Singapore 487372, Singapore}

\begin{abstract}
We study the possibility of  triply-degenerate points (TPs) that can be stabilized in spinless crystalline systems. Based on an exhaustive search over all 230 space groups, we find that the spinless TPs can exist at both high-symmetry points and high-symmetry paths, and they may have either linear or quadratic dispersions. For TPs located at high-symmetry points,  they all share a common minimal set of symmetries, which is the point group $T$. The TP protected solely by the $T$ group is chiral and has a Chern number of $\pm2$. By incorporating additional symmetries, this TP can evolve into  chiral pseudospin-1 point, linear TP without chirality, or quadratic contact TP. For accidental TPs  residing  on a high-symmetry path, they are not chiral but can have either linear or quadratic dispersions in the plane normal to the path. We further construct effective $k\cdot p$ models and minimal lattice models for characterizing these TPs.  Distinguished phenomena for the chiral TPs are discussed, including the extensive surface Fermi arcs and the chiral Landau bands.

\end{abstract}
\maketitle

\section{\label{intro}Introduction}

Physical properties of metals are largely determined by the electronic quasiparticles around the Fermi level. In usual cases, these electrons can be well approximated as free particles characterized by a renormalized effective mass~\cite{Ashcroft1976}. However, there are also special cases in which this treatment is no longer valid, namely, when certain band degeneracy points (BDPs) are located at or close to the Fermi level. These points may change the fundamental character of the low-energy quasiparticles~\cite{Bansil2016,Chiu2016,Armitage2018}. First, the particles acquire a pseudospin degree of freedom corresponding to the degeneracy of the BDP. Second, the energy dispersion of the particles could be different and is determined by the character of the BDP. Third, the particles may even acquire chirality when the BDP carries a chiral topological charge. As prominent examples, Weyl and Dirac particles emerge around twofold degenerate Weyl and fourfold degenerate Dirac points~\cite{Wan2011,Young2012,Wang2012,Wang2013,Dai2016,Armitage2018}, respectively. They are massless with linear dispersions. And the Weyl particles have a definite chirality, corresponding to the Chern number of the Weyl point.

In condensed matters, the BDPs and the associated emergent quasiparticles have a much richer variety compared to the high energy physics, because the symmetry requirement,  i.e., the space group (SG) symmetry, is much reduced from the Poincar$\acute{e}$ group for elementary particles~\cite{Bradlyn2016}. Notably, a kind of three-component particles, emerging at triply-degenerate points (TPs), was proposed and attracted great interest. It was predicted in  WC-family materials~\cite{Zhu2016}, $\theta$-TaN~\cite{Weng2016}, and many other systems~\cite{Weng2016a,Sun2017a,Chang2017,Zhang2017a,Sun2017,Yang2017,Wang2017,Xia2017,Guo2018,Jin2019}. TPs in some of these proposals have been successfully demonstrated in experiment~\cite{Lv2017,Ma2018,Gao2018,Rao2019,Yuan2019}. Physically, the TP particles may be regarded as intermediate between two-component Weyl and four-component Dirac particles. It was shown that they may lead to interesting surface states, unusual transport property, and topological phase transitions~\cite{Bradlyn2016,Weng2016,Zhu2016,Hu2018}.

So far, most studies on TPs are in systems with spin-orbit coupling (SOC). For example, the material examples mentioned above have sizable SOC. In Ref.~\cite{Bradlyn2016}, Bradlyn \emph{et al.} classified possible TPs at high-symmetry points of the Brillouin zone (BZ) for systems with SOC. On the other hand, it was noted that TPs may also appear in spinless systems. For example, linear and special quadratic TPs were reported in three-dimensional (3D) honeycomb carbon~\cite{Hu2019} and H-boron~\cite{Gao2019}, where the SOC can be neglected. In addition, the discussion of topological spinless systems also applies to the huge fields of bosonic and even classical periodic systems, which are currently under rapid development~\cite{Lu2014,Yang2015,Mittal2019,Xue2020,Imhof2018,Yu2020,Huber2016,Prodan2009}. Indeed, a recent work by Yang \emph{et al.}~\cite{Yang2019} demonstrated a special chiral TP in a phononic crystal. Therefore, it is an important task to have a systematic analysis of TPs in spinless systems. The study will help us address the following open questions: \emph{What are all possible kinds of TPs in spinless systems? What are their symmetry requirements? What are the properties of the emergent TP particles?}

{
\global\long\def\arraystretch{1.4}%

\begin{table*}
\caption{List of TPs at high-symmetry points identified in 230 SGs for spinless systems. Here, $|\mathcal{C}|$ is the Chern number of the point. }
\label{thsp}
\begin{tabular}{llllll}
\hline
\hline
Order & Minimal symmetry   & Adding symmetry & SG and Location & $|{\cal C}|$ & notation\\
\hline
Linear & $C_{3,111}$, $C_{2z}$, $C_{2y}$ &  & 197, $P$ & 2 & C-2 TP\\

 &  & ${\cal T}C_{2;110}$ & 211, $P$  & 2 & \\

 &  & ${\cal T}$ & 195-199, $\Gamma$; 195, $R$;
197, 199, $H$  & 2 & \\

 &  & $C_{2,110}$, ${\cal T}$ & 207-214, $\Gamma$; 207-208, $R$;
211, 214, $H$ & 2 & \\

 &  & ${\cal P}{\cal T}$ & 204, $P$ & - & TP\\

 &  & $M_{110}$ & 217, $P$ & -   & \\

 &  &$M_{110}$, ${\cal P}{\cal T}$ & 229, $P$ & -  & \\
\hline
Quadratic & & ${\cal P}$, ${\cal T}$  & 200-206, $\Gamma$; 200-201 $R$; 204, 206, $H$ & 0 & QCTP \\

 &  & $M_{110}$,  ${\cal T}$ & 215-220, $\Gamma$; 215, $R$; 217, $H$ & 0  & \\

 &  & ${\cal P}$, $M_{110}$,  ${\cal T}$ & 221-230, $\Gamma$; 221, 224, $R$; 229, $H$ & 0  & \\

\hline
\hline
\end{tabular}

\end{table*}

}

In this work, we undertake this task and answer the above questions. We perform an exhaustive search of all possible TPs in the 230 SGs for 3D spinless systems with time reversal symmetry. The results are summarized in Tables~\ref{thsp} and \ref{tab:thsl}. Our key findings are the following. (i) TPs can be classified into two large classes: those at high-symmetry points of the BZ and those on high-symmetry lines. All these points are isolated, meaning that there are no triply-degenerate nodal lines or surfaces. (ii) For each class, there are two subclasses according to the dispersion of the TP: linear or quadratic. In other words, we find that the leading order in dispersion for a TP cannot be higher than two.
(iii) Chiral TPs only appear for linear TPs at high-symmetry points. Notably, they may reside at both time-reversal-invariant momentum (TRIM) points and non-TRIM points. This is in contrast to chiral TPs in spinful systems, which cannot appear at TRIM points and must require certain nonsymmorphic symmetry. We present the symmetry conditions and the $k\cdot p$ effective model for each kind of TPs. For the most interesting cases with chiral TPs and quadratic TPs, we also construct lattice models to demonstrate their existence.
The manifestation of the chiral TPs in topological surface states and Landau spectra is discussed.

Our work provides a comprehensive view of TPs in spinless systems. The results will be useful for searching and studying TP particles in real materials as well as designed artificial structures.

\section{\label{Appro} Approach }

To obtain a complete classification of TPs in spinless systems, we scan through all the irreducible representations (IRRs) of the little group at high-symmetry points and lines in the BZ for each of the 230 SGs (clearly, TPs cannot occur at generic $k$ points). For spinless systems, these IRRs correspond to the single-valued representations, which have been tabulated in standard references~\cite{Bradley2010}.

At high symmetry points, a TP corresponds to a 3D IRR of the little group. On a high-symmetry line, a TP corresponds to a crossing between bands with
2D and 1D IRRs. We investigate all these possibilities with the knowledge of the IRRs.

For each identified TP, we characterize it by constructing the $k\cdot p$ effective model $H_{\text{eff}}$ from the symmetry constraints
\begin{align}
D(\mathcal{G}_{i})H_{\text{eff}}(\mathcal{G}_{i}^{-1}\bm{k})D^{-1}(\mathcal{G}_{i})=H_{\text{eff}}(\boldsymbol{k}),
\label{constrain}
\end{align}
where ${\cal G}_{i}$ is the $i$-th generator of the little group at TP, and $D(\mathcal{G}_{i})$ stands for the matrix representation of $\mathcal{G}_{i}$. The dispersion and the chiral charge can be obtained from this effective model. The similar approach is also used for constructing the lattice models for a few representative SGs.

In the following, we shall discuss the obtained results in Tables~\ref{thsp} and \ref{tab:thsl}.

\section{\label{hsp} at high-symmetry points}

TPs at high-symmetry points correspond to the 3D IRRs of the little group at the point. We find that all these TPs share a common minimal set of symmetries. In Sec.~\ref{Msymm}, we shall first discuss this minimal set and present the most general model. Then, in Sec.~\ref{SymmAsc}, we shall add additional symmetries to this minimal set (denoted as ``symmetry ascending") and investigate the impact on the character of the TP.

\subsection{\label{Msymm} Minimal Symmetry }

Let $G$ be the little group at a high-symmetry point, a TP is associated with the 3D single-valued IRR of $G$. The minimal symmetry would then be the simplest $G$ that possesses a 3D IRR. We find that this is given by the point group $T$.

Group $T$ is generated by three elements: a twofold rotation $C_{2y}$, a twofold rotation $C_{2y}$, and a threefold rotation $C_{3;111}$. Here, the rotation axis for $C_{3;111}$ is along the $[111]$ direction. The 3D IRR for $T$ corresponds to the basis states of $\Psi=(p_x,p_y,p_z)$. Under this basis, the representations of the three generators are given by
\begin{align}\nonumber
D(C_{3;111})=&\begin{bmatrix}
0 & 0 & 1\\
1 & 0 & 0\\
0 & 1 & 0
\end{bmatrix},~D(C_{2z})=\begin{bmatrix}
-1 & 0 & 0\\
0 & -1 & 0\\
0 & 0 & 1
\end{bmatrix},\\~
D(C_{2y})=&\begin{bmatrix}
-1 & 0 & 0 \\
0 & 1 & 0 \\
0 & 0 & -1
\end{bmatrix}.\label{REP}
\end{align}

The effective model for the corresponding TP can be derived according to Eqs.~(\ref{constrain}) and (\ref{REP}). The obtained model expanded to the first order in $k$ can be expressed in a compact form as
\begin{eqnarray} \label{ess-mim}
 H &=& c_1 H_\text{chiral} +c_2 H_\text{achiral},
\end{eqnarray}
with
\begin{eqnarray} \label{chiral}
 H_\text{chiral} &=& k_x\Lambda_{7}-k_y\Lambda_{5}+k_z\Lambda_{2},
\end{eqnarray}
and
\begin{eqnarray} \label{achiral}
 H_\text{achiral} &=& k_{x}\Lambda_{6}+k_{y}\Lambda_{4}+k_z\Lambda_{1}.
\end{eqnarray}
Here, the energy and the momentum are measured from the TP, $c_1$ and $c_2$ are real parameters, and $\Lambda_{i}$ are the $3\times 3$ Gell-Mann matrices (see Appendix~\ref{gell} for their concrete forms). 
If we only have the first term in Eq.~(\ref{ess-mim}), the TP would feature a Chern number of $\pm 2$. In contrast, if  the TP  only has  the second term in Eq.~(\ref{ess-mim}), it does not have a well defined Chern number, as one crossing band is doubly  degenerate along certain high-symmetry paths. This explains the meaning of the subscripts.
In the general case, i.e., with group $T$, both terms should be present, and the TP is chiral and has Chern number $\pm 2$. Such TPs are termed as charge-2 TPs. Interestingly, we note that in spinful systems, the existence of chiral TPs would require nonsymmorphic symmetries~\cite{Bradlyn2016}. In comparison, for spinless systems, symmorphic symmetries are sufficient to stabilize a chiral TP.

This minimal symmetry case is met at point $P$ for the SG~197, as shown in Table~\ref{thsp}. When additional symmetries are added, some of the terms in Eq.~(\ref{ess-mim}) could be eliminated, and the TP may be transformed to other types. The result of such symmetry ascending process will be discussed in the following.

\subsection{\label{SymmAsc} Symmetry Ascending}
\subsubsection{\label{C-2 TP} Charge-2 TP}

As we demonstrated above, the minimal symmetry condition gives a charge-2 TP described by Eq.~(\ref{ess-mim}).
By adding additional symmetries, such as the time reversal symmetry $\cal{T}$, twofold rotation $C_{2;110}$ along the $[110]$ direction, or their combination $\mathcal{T} C_{2;110}$, we find that the TP will remain a charge-2 TP with linear dispersion in all direction (see Table~\ref{thsp}), but its effective model is greatly simplified.

Consider the TP at the $\Gamma$ point of SG~195 as an example. The little cogroup is $G= T\otimes \{{\cal{T}}, E\}$ with $E$ the identity element.  The matrix representation of $\cal{T}$ under the  basis $\Psi$ is
\begin{eqnarray}
 D({\cal{T}}) &=& \mathbb{I}_3 {\cal{K}},
\end{eqnarray}
with $\cal{K}$ the complex conjugation and $\mathbb{I}_3$ the $3\times 3$ identity matrix. Clearly, the added $\cal{T}$ symmetry eliminates the second term in Eq.~(\ref{ess-mim}), since the Gell-Mann matrices involved in $H_\text{achiral}$ are purely real.
Then, the low-energy model for this TP takes a very simple form of
\begin{eqnarray} \label{spin-1}
 H_{195}^{\Gamma} &=& c H_\text{chiral}.
\end{eqnarray}
Notice that the Gell-Mann matrices in $H_\text{chiral}$ satisfy the algebra of angular momentum operators $\left[S_i,S_j\right]=i\epsilon_{ijk}S_k$,  if we set $S_x=\Lambda_{7}$, $S_y=-\Lambda_5$, and $S_z=\Lambda_2$. Hence, the charge-2 TP here corresponds to the chiral pseudospin-1 particles~\cite{Bradlyn2016}.
It was shown that such particles can exhibit remarkable effects such as super Klein tunneling~\cite{Shen2010}, supercollimation~\cite{Fang2016}, and super Andreev reflection~\cite{Feng2020}. The chiral TP reported in the recent experiment on a phononic crystal (with SG~198) also belongs to this category~\cite{Yang2019}.

\subsubsection{Linear achiral TP}
With the addition of $M_{110}$, the combination of $\cal{T}$ and inversion symmetry  ${\cal{P}}$, or both $M_{110}$ and ${\cal{PT}}$ (see Table \ref{thsp}), the original charge-2 TP in (\ref{ess-mim}) would transform into a TP without a finite Chern number. This can be easily understood, since the ${\cal{PT}}$ symmetry suppresses the Berry curvature field and any monopole topological charge cannot reside on a mirror plane.

Consider the triple point at the $P$ point of SG 204, which has ${\cal{PT}}$ in addition to the minimum symmetry $T$. The matrix representation of $\cal{PT}$ under the  basis $\Psi$ can be written as
\begin{eqnarray}
 D({\cal{PT}}) &=& -\mathbb{I}_3 {\cal{K}}.
\end{eqnarray}
This symmetry eliminates the first term in (\ref{ess-mim}), as the Gell-Mann matrices in it are purely imaginary. Hence, we obtain
\begin{align}\label{ham204p}
H_{204}^{P} =  c  H_\text{achiral}.
\end{align}
This TP has linear dispersion, vanishing Berry curvature, and is not chiral.
Moreover, a detailed analysis shows that two of the three bands must be degenerate along the $P$-$H$ and $P$-$\Gamma$ paths, due to the $C_3$ symmetry.

\begin{table*}
\caption{List of TPs on high-symmetry lines identified in 230 SGs for spinless systems. It should be noted that they are categorized according to their local symmetries. For instance, $\{C_{4z}, M_{y}\}$ and $\{C_{4y}, M_{x}\}$ are the same.
}
\label{tab:thsl}
\begin{tabular}{clll}
\hline
\hline
Order & Minimal Symmetry & Adding Symmetry & SG and Location \\
\hline

Linear & $C_{2z}$, $S_{4z}\mathcal{T}$ &  & 81-82, $\Lambda$; 81-82, $V$\\

 &  & $M_{y}$ & 111-122, $\Lambda$; 111-112, 115-121, $V$; 122, $V$; 215, 218, $T$; {\color{black} 215-220, $\Delta$}\\

 & $C_{3z}$, $S_{3z}\mathcal{T}$ &  & 174, $\Delta$\\

 & $C_{3z}$, ${\cal P}{\cal T}$ &  & 147, $\Delta$; 148, $\Lambda$; 147-148, 164-165, 175-176, $P$; 200-206, $\Lambda$; 204, 206, $F$\\

 & $C_{4z}$, ${\cal P}{\cal T}$ &  & 83-88, $\Lambda$; 83-88,$V$\\

 & $C_{3z}$, $M_{x}$ &  & 156-159, $\Delta$; 160-161, $\Lambda$; 157, 159-161, 183-186, 189-190, $P$\\

 &  & ${\cal P}{\cal T}$ & 162-165, $\Delta$; 166-167, $\Lambda$; 162-163, 166-167, 191-194, $P$\\

 &  & $S_{3z}\mathcal{T}$ & 187-190, $\Delta$\\

 & $C_{3;111}$, {\color{black} $M_{110}$} &  & 215-220, $\Lambda$; {\color{black} 217, 220, $F$}\\

 &  & ${\cal P}{\cal T}$ & 221-230, $\Lambda$; {\color{black} 229-230, $F$} \\

 & $C_{4z}$, $M_{y}$ &  & 99-110, $\Lambda$; 99, 101, 103, 105, 107-108, $V$\\

 &  & {\color{black} $C_{2z}$} & 100, 102, 104, 106, 109-110, $V$\\

 &  & ${\cal P}{\cal T}$ & 123-142, $\Lambda$; 123-124, 131-132, 139-142,$V$; 221, 223, $T$; {\color{black} 221-230, $\Delta$}\\

 &  & {\color{black} $C_{2z}$}, ${\cal P}{\cal T}$ & 125-126, 133-134, $V$; 222, $T$\\

 & $C_{2z}$, $M_{x}$, $M_{110}$ &  & \\

 &  & ${\cal P}{\cal T}$ & 224, $T$\\

 \hline
Quadratic & $C_{6z}$, ${\cal P}{\cal T}$ &  & 175-176, $\Delta$\\

 & $C_{6z}$, $M_{x}$ &  & 183-186, $\Delta$\\

 &   & ${\cal P}{\cal T}$ & 191-194, $\Delta$\\
\hline
\hline
\end{tabular}
\end{table*}

\subsubsection{Quadratic contact TP}

Apart from the linear TPs discussed above, we also find TPs exhibiting quadratic energy splitting along all directions in momentum space, as shown in Table \ref{thsp}.
This is possible when the additional symmetry is ${\cal{P}}$ or  ${\cal{T}}M_{110}$.

Let us first consider the consequence of adding ${\cal{P}}$. This applies to the $\Gamma$ point of SG~204. Then, the  point group $T$ is transformed into $T_h$.  Since all the ${\cal{P}}$-invariant points are also TRIM points, the $\mathcal{T}$ symmetry must also be present.
The matrix representation of ${\cal{P}}$ under the basis state $\Psi$ reads
\begin{eqnarray}
 D({\cal{P}}) &=& -\mathbb{I}_3.
\end{eqnarray}
According to the constraint
\begin{align} \label{con-QCTP}
D({\cal{P}})H_{\text{eff}}(-\bm{k})D^{-1}({\cal{P}})=H_{\text{eff}}(\boldsymbol{k}),
\end{align}
all the odd-order terms in $k$ must be excluded. Therefore, the leading order becomes $k$ quadratic.
Expanded up to the $k$-quadratic order, we obtain
\begin{align}\nonumber
H_{204}^{\Gamma}&=c_1k^2\mathbb{I}_{3}+c_2(\Lambda_{1}k_xk_y+\Lambda_{4}k_xk_z+\Lambda_{6}k_yk_z)\\ \nonumber
+ &(\sqrt{3} c_3 \Lambda_3 + (c_3+2c_4)\Lambda_8) k_x^2 - (\sqrt{3}(c_4+c_3)\Lambda_3 \\
+ &(c_4-c_3)\Lambda_8) k_y^2 + (\sqrt{3} c_4 \Lambda_3 -(2c_3+c_4)\Lambda_8 )k_z^2.
\label{kp204}
\end{align}
Clearly, this TP has a quadratic energy splitting along all directions in the momentum space.
We term this kind of TP as quadratic contact TP (QCTP) to indicate that the bands ``contact" rather than ``cross" each other~\cite{Zhu2018}.

The QCTP can also be protected by  the $T_d$ group together with ${\cal{T}}$, corresponding to adding a vertical mirror ${\cal{M}}_{110}$ and ${\cal{T}}$ to $T$. This applies for example to the $\Gamma$ point for SG~215.  In such a case, the low-energy effective model reads
\begin{align}\nonumber
H_{215}^{\Gamma}&=c_1k^2\mathbb{I}_{3}+ c_2(k_x k_y \Lambda_1 +  k_x k_z \Lambda_4+ k_y k_z \Lambda_6) \\
&+ c_3\left[\sqrt{3}(k_x^2-k_y^2)\Lambda_3 +  (k_x^2 + k_y^2 -2k_z^2)\Lambda_8\right].
\label{kp215}
\end{align}

Finally, with the addition of symmetry ${\cal{M}}_{110}$ (${\cal{P}}$), the point group $T_h$  ($T_d$) will be further transformed into the $O_h$ group, which has four different 3D IRRs.
We find that the TPs protected by these IRRs of $O_h$  are also QCTPs.
And one notes that since the QCTPs have either ${\cal{PT}}$ or  $M_{110}$  symmetry, the Chern numbers for all the three bands of a QCTP must be zero.

\section{\label{hsl} On high-symmetry lines}

TPs on high-symmetry lines are formed by the crossing between a non-degenerate band and a doubly degenerate band. In this sense, these TPs belong to the accidental band crossings.
The collection of such TPs from our systematic search is listed  in Table~\ref{tab:thsl}. We note that first, while the TPs at high-symmetry points are only available in cubic crystal systems, the existence of TPs on high-symmetry lines is more extensive. Besides cubic systems, they also appear in tetragonal, trigonal and hexagonal systems. Second, in addition to the linear TPs, we also have quadratic TPs on high-symmetry lines. This is distinct from spinful systems, where quadratic TPs cannot exist on high-symmetry lines~\cite{Bradlyn2016,Zhu2016}.
Third, all TPs on high symmetry lines do not have a well defined Chern number. This can be easily understood by noting that one crossing band is doubly degenerate along the high-symmetry path. Last,
different from the QCTPs at high symmetry points which have quadratic dispersion along all directions, the quadratic TPs on high symmetry lines have linear dispersion along the line and quadratic dispersion in the plane normal to the line.

\begin{table}[t]
\centering
\caption{IRRs for Quadratic TPs on high-symmetry lines. Quadratic TPs can be found only on sixfold axis in SGs 175-176, 183-186, and 191-194, which also require specific IRRs provided here. In the table, PG and AG denote point group and abstract group, respectively. The notations of AG and the corresponding IRRs are adopted from Ref.~\cite{Bradley2010}.}
\begin{tabular} {cccl}
\hline
\hline
  SGs  &  PG & AG  & IRRs \\
\hline
$175{-}176$ &$C_{6}$ & $G_{6}^1$ & $\{R_1,R_3\oplus R_5\},\{R_2\oplus R_6,R_4\}$ \\
$183{-}186$ &$C_{6v}$ & $G_{12}^3$ & $\{R_1,R_5\},\{R_2,R_5\},\{R_3,R_6\},\{R_4,R_6\}$  \\
$191{-}194$ &$C_{6v}$ & $G_{12}^3$ & $\{R_1,R_5\},\{R_2,R_5\},\{R_3,R_6\},\{R_4,R_6\}$  \\
\hline
\hline
\label{dqhsl}
\end{tabular}
\end{table}

\subsection{\label{Msymml} Minimal Symmetry}
There are three different types of minimal symmetry conditions
for protecting TPs on a high-symmetry line.
The first one is the group $C_{nv}$ (with $n=3, 4, 6$), which has both 1D and 2D single-valued IRRs.
The second is the three-fold rotation $C_{3z}$ together with $\cal{PT}$ or $S_{3z}\cal{T}$, as the pair of conjugated 1D single-valued IRR of $C_{3z}$ are bound together into a 2D co-representation by  $\cal{PT}$ or $S_{3z}\cal{T}$. Here, $S_{3z}$ is the threefold roto-reflection along $z$. The last case is a two-fold rotation $C_{2z}$ together with  $S_{4z}\cal{T}$. The bands on the high-symmetry line along $z$ can be labeled by the  eigenvalues of $C_{2z}$, which are $\pm 1$. Correspondingly, we have
${(S_{4z}{\cal{T}})}^2=C_{2z}=\pm 1$, so the bands with $C_{2z}=1$ are non-degenerate, while the bands with $C_{2z}=-1$ must be doubly degenerate due to the Kramers-like constraint   $(S_{4z}{\cal{T}})^2=-1$.
A crossing between a non-degenerate band and a doubly degenerate band will then form an accidental TP.

With symmetry ascending, the accidental TPs protected by the minimal symmetry conditions may be transformed  into other types of BDPs. All the cases giving TPs are listed in Table \ref{tab:thsl}.

From the Table \ref{tab:thsl}, one finds most of the accidental TPs show linear energy splitting.
There do exist several TPs exhibiting quadratic energy splitting in the plane normal to the high-symmetry line.
While the linear accidental TPs have been well studied in both spinless and  spinful  systems in previous works, especially in the WC-family materials~\cite{Weng2016,Zhu2016,Weng2016a,Lv2017,Ma2018}, the report of quadratic TPs is very limited. In a recent work~\cite{Gao2019}, the quadratic TP was found in H-boron, with SG~194.
In the following section, we shall give a detailed discussion of the quadratic TPs.

\subsection{\label{QTP}Quadratic TP}

As shown in Table \ref{tab:thsl}, Quadratic TPs only occur on the high-symmetry line ($\Delta$) with sixfold rotation $C_{6z}$ in SGs 175-176, 183-186, and 191-194, where the line $\Delta$ also has $\mathcal{PT}$, $\mathcal{M}_x$ and both $\mathcal{PT}$ and $\mathcal{M}_x$  symmetry, respectively.
The appearance of quadratic TPs also requires the crossing between bands with a particular pair of IRRs, which are presented in Table~\ref{dqhsl}.

Let us take the quadratic TP formed by the $R_4$ and $R_6$ IRRs on the  $\Delta$ line in SG~191 as an example. The matrix representations for the symmetries can be taken as
\begin{align}
D(C_6)=\begin{bmatrix}
-1 & 0 & 0 \\
0 & \frac{1}{2} & -\frac{\sqrt{3}}{2} \\
0 & \frac{\sqrt{3}}{2} & \frac{1}{2}
\end{bmatrix},~D(M_x)=\begin{bmatrix}
-1 & 0 & 0 \\
0 & 1 & 0 \\
0 & 0 & -1
\end{bmatrix},
\end{align}
and $D(\mathcal{PT})=\mathbb{I}_3\mathcal{K}$.
Clearly, we have
\begin{equation}
  D(C_6)^3=D(C_{2z})=-\mathbb{I}_3,
\end{equation}
and the twofold rotation requires that
\begin{align} \label{con-QCTP}
D(C_{2z})H_{\text{eff}}(-k_x,-k_y,k_z)D^{-1}(C_{2z})=H_{\text{eff}}(\bm k),
\end{align}
which eliminates all the terms with odd orders in $k_x$ and $k_y$.  Therefore, the leading order for dispersion in the $k_x$-$k_y$ plane becomes quadratic.
Expanded to the leading order in each direction, the  effective model for this quadratic TP can be obtained as
\begin{align}\nonumber
H_{191}^{\Delta}&=\left[c_1k_z+c_2k_z^2+c_3(k_x^2+k_y^2)\right]\mathbb{I}_3 + c_4(k_x^2-k_y^2)\Lambda_4  \\\nonumber
 + &(2c_4\Lambda_1 + c_5\Lambda_6) k_x k_y  + (3\Lambda_3+\sqrt{3}\Lambda_8)(c_6k_z+c_7k_z^2) \\\nonumber
 + &[\sqrt{3}c_8\Lambda_3+(2c_9-c_8)\Lambda_8]k_x^2 \\
 + &[\sqrt{3}c_9\Lambda_3+(2c_8-c_9)\Lambda_8]k_y^2.
\label{kp191}
\end{align}

\begin{figure*}[t]
	\centering
	\includegraphics[angle=0, width=0.95\textwidth]{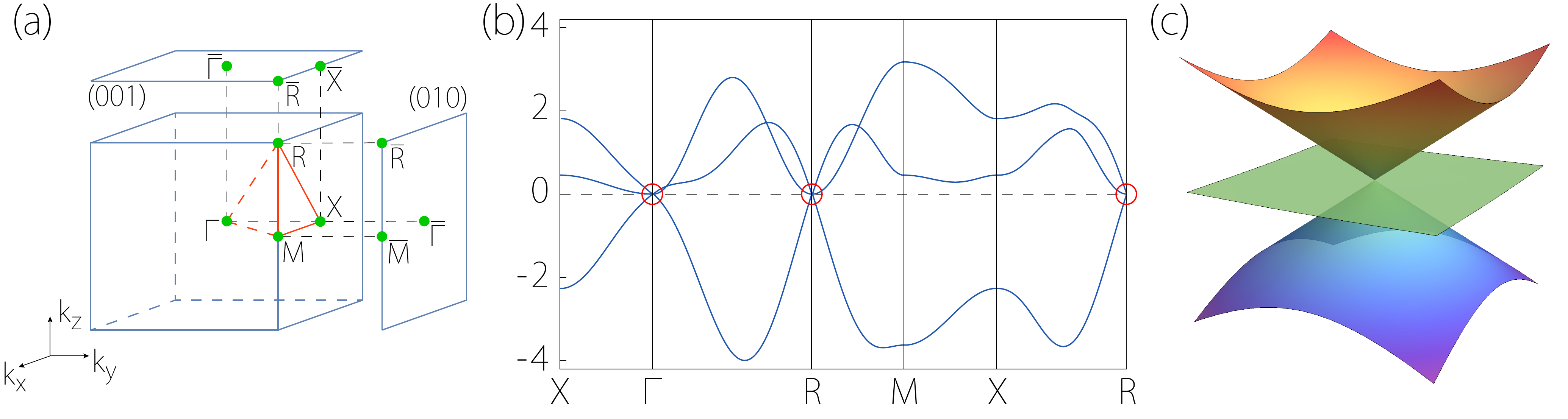}
	\caption{(a) Brillouin zone of SG 195. (b) Band structure of the tight-binding model~(\ref{tb195}) with SG~195. The parameters in the model are taken as $c_1=-1.4,~c_2=1.4,~c_3=0.8,~c_4=0.6$, and $c_5=-0.4$. The red circles denote the two chiral TPs at $\Gamma$ and R points. (c) Linear energy dispersion in the vicinity of the chiral TP.\label{fig:SG195}}
\end{figure*}

\section{Lattice models}
In this section, we present lattice models for three most interesting cases, namely the charge-2 TP, the QCTP, and the quadratic TP, in order to explicitly demonstrate their existence. These models will also serve as a good starting point for studying their physical properties.

\subsection{Lattice Model for Charge-2 TP}

Let us first consider a lattice model with SG~195, which contains the charge-2 TP. We take a simple cubic lattice, and assign three active orbitals  \{$p_x,p_y,p_z$\} at the lattice sites (corresponding to the $1a$ Wyckoff position). We find that the following lattice model satisfies all the symmetry constraints of SG~195:
\begin{align}\nonumber
H_{195}^\text{TB} &=  c_1(\Lambda_2\sin k_z-\Lambda_5\sin k_y+\Lambda_7\sin k_x)  \\ \nonumber
+ &c_2(\Lambda_1 \sin k_x \sin k_y + \Lambda_4 \sin k_x \sin k_z + \Lambda_6 \sin k_y \sin k_z)\\ \nonumber
+ &c_3(\Lambda_2 \cos k_y \sin k_z - \Lambda_5 \cos k_x \sin k_y + \Lambda_7 \cos k_z \sin k_x) \\\nonumber
+ &( \sqrt{3} \Lambda_3 - \Lambda_8 )(c_4 + c_5 \cos k_z) \cos k_x \\\nonumber
- &( \sqrt{3} \Lambda_3 + \Lambda_8 )(c_4 + c_5 \cos k_y) \cos k_z \\
+ &2\Lambda_8(c_4 + c_5 \cos k_x)  \cos k_y.
\label{tb195}
\end{align}
The calculated band structure of this model is plotted in Fig.~\ref{fig:SG195}(b). One can clearly observe two triply degenerate points at $\Gamma$ and $R$ points. These points have linear energy splitting, as illustrated in Fig.~\ref{fig:SG195}(c). Each of them is described by the model in Eq.~(\ref{spin-1}) (here $R$ point has the same symmetry as $\Gamma$). For the parameters taken in Fig.~\ref{fig:SG195}(b), we verify that the TP at $\Gamma$ has a Chern number of -2, whereas the TP at $R$ has a Chern number of 2.

It should be pointed out that in the current case, the two charge-2 TPs can be the only BDPs around the Fermi level (if considering a spinless fermionic systems). This is in contrast to the spinful systems, where a charge-2 TP must coexist with other kinds of BDPs, such as those at TRIM point~\cite{Bradlyn2016}.

\subsection{Lattice Model for QCTP}

\begin{figure}[t]
	\centering
	\includegraphics[angle=0, width=0.48\textwidth]{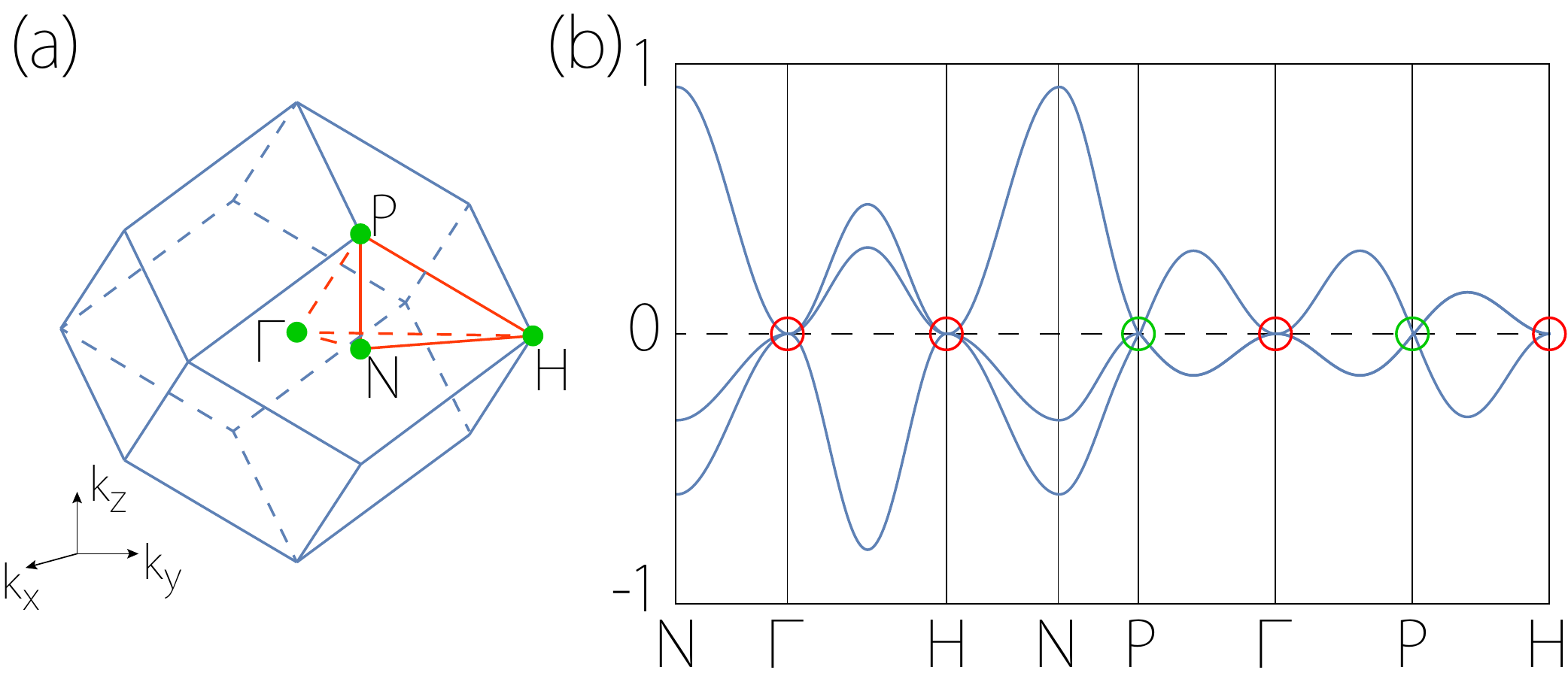}
	\caption{(a) Brillouin zone for SG 204. (b) Band structure of the tight-binding model~(\ref{tb204}) with SG~204. The model parameters are taken as $c_1=0.4,~c_2=-0.2,$ and $c_3=-0.02$. The red circles denote the QCTPs at $\Gamma$ and $H$ points. The green circle denotes the achiral TP at $P$ with linear dispersion in all directions.  }
	\label{tdpsg204}
\end{figure}

For the QCTP, we consider a lattice model with SG~204. Again, we take a simple cubic lattice, with three active orbits \{$p_x,p_y,p_z$\} on each lattice site. The model we obtain is given by
\begin{align}\nonumber
H_{204}^\text{TB} &= c_1(\Lambda_1\cos k_z \sin k_x \sin k_y +\Lambda_4\cos k_y \sin k_x \sin k_z \\\nonumber
+& \Lambda_6 \cos k_x \sin k_y \sin k_z) + [ c_2(\sqrt{3} \Lambda_3-\Lambda_8) \\\nonumber
-& 2c_3\Lambda_8 ] \cos 2k_x + [ 2c_2\Lambda_8  +c_3(\sqrt{3} \Lambda_3+\Lambda_8) ]\cos 2k_y \\
+& [c_3(\Lambda_8-\sqrt{3}\Lambda_3) - c_2 (\Lambda_8+\sqrt{3}\Lambda_3)] \cos 2k_z.
\label{tb204}
\end{align}
The calculated band structure is plotted in Fig.~\ref{tdpsg204}(b). One can clearly observe two QCTPs at $\Gamma$ and $H$ points. We confirm that the leading order energy splitting around these points are of quadratic order. Meanwhile, there is another linear achiral TP at the $P$ point, which is consistent with results in Table~\ref{thsp}.

\subsection{Lattice model for Quadratic TP}
For the quadratic TP on a high-symmetry line, we consider a lattice model with SG~191. We take a trigonal lattice, with three active orbitals \{$d_{x^2-y^2},d_{xy},d_{z^2}$\} on each lattice site (1$a$ Wyckoff position). The lattice model that satisfies all symmetry constraints can be taken as
\begin{align}\nonumber
H_{191}^\text{TB} &=c_1  \bigg[ \Lambda_1  \bigg( \cos k_y - \cos \frac{\sqrt{3}k_x}{2}\cos\frac{k_y}{2}  \bigg) \\\nonumber
& + \sqrt{3} \Lambda_4 \sin \frac{\sqrt{3}k_x}{2} \sin \frac{k_y}{2}  \bigg] + (\sqrt{3}\Lambda_3+\Lambda_8) \bigg[ c_2\cos k_z \\
& + c_{3} \bigg( \cos k_y + 2\cos \frac{\sqrt{3}k_x}{2} \cos \frac{k_y}{2} \bigg) \bigg].
\label{tb191}
\end{align}

\begin{figure}[t]
	\centering
	\includegraphics[angle=0, width=0.49\textwidth]{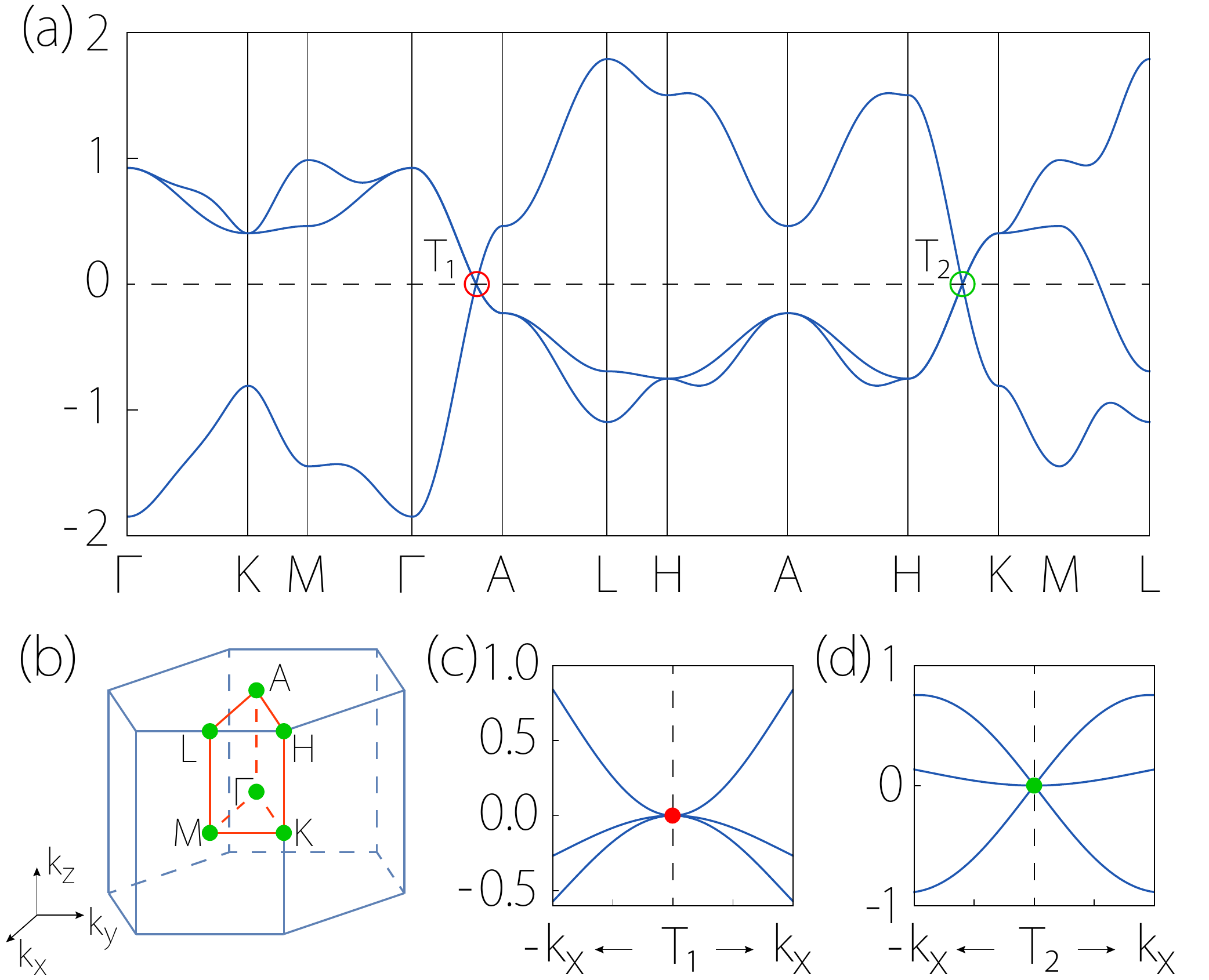}
	\caption{(a) Band structure of the tight-binding model~(\ref{tb191}) with SG 191. The model parameters are taken as $c_1=c_2=-0.5$, and $c_3=-0.1$. The red circle indicates the Quadratic TP on the $\Gamma$-$A$ $(\Delta)$ path. The green circle indicates a linear TP on the $H$-$K$ $(P)$ path. (b) Brillouin zone of SG 191. (c) and (d) show the energy dispersion around the quadratic TP and the linear TP, respectively.}
	\label{tdpsg191}
\end{figure}

The calculated band structure can be found in Fig.~\ref{tdpsg191}. One observes that two TPs appear on the $\Gamma$-$A$ and the $H$-$K$ paths,
with linear band crossing along $k_z$. By investigating the band dispersion around these points in the plane normal to $k_z$, we confirm that the TP on $\Gamma$-$A$ is a quadratic TP [see Fig.~\ref{tdpsg191}(c)], whereas that on $H$-$K$ is a linear TP [see Fig.~\ref{tdpsg191}(d)], consistent with the results in
Table~\ref{tab:thsl}.

\section{\label{chiralTP} Property of chiral TP}

Due to the nonzero Chern number, the charge-2 TPs can exhibit many interesting physical properties. In this section, we will highlight two examples. One is the topologically protected long surface Fermi arcs, and the other is the chiral Landau bands under a magnetic field.

\subsection{\label{sfarc} Extensive Fermi Arcs}

\begin{figure}[t]
	\centering
	\includegraphics[angle=0, width=0.48\textwidth]{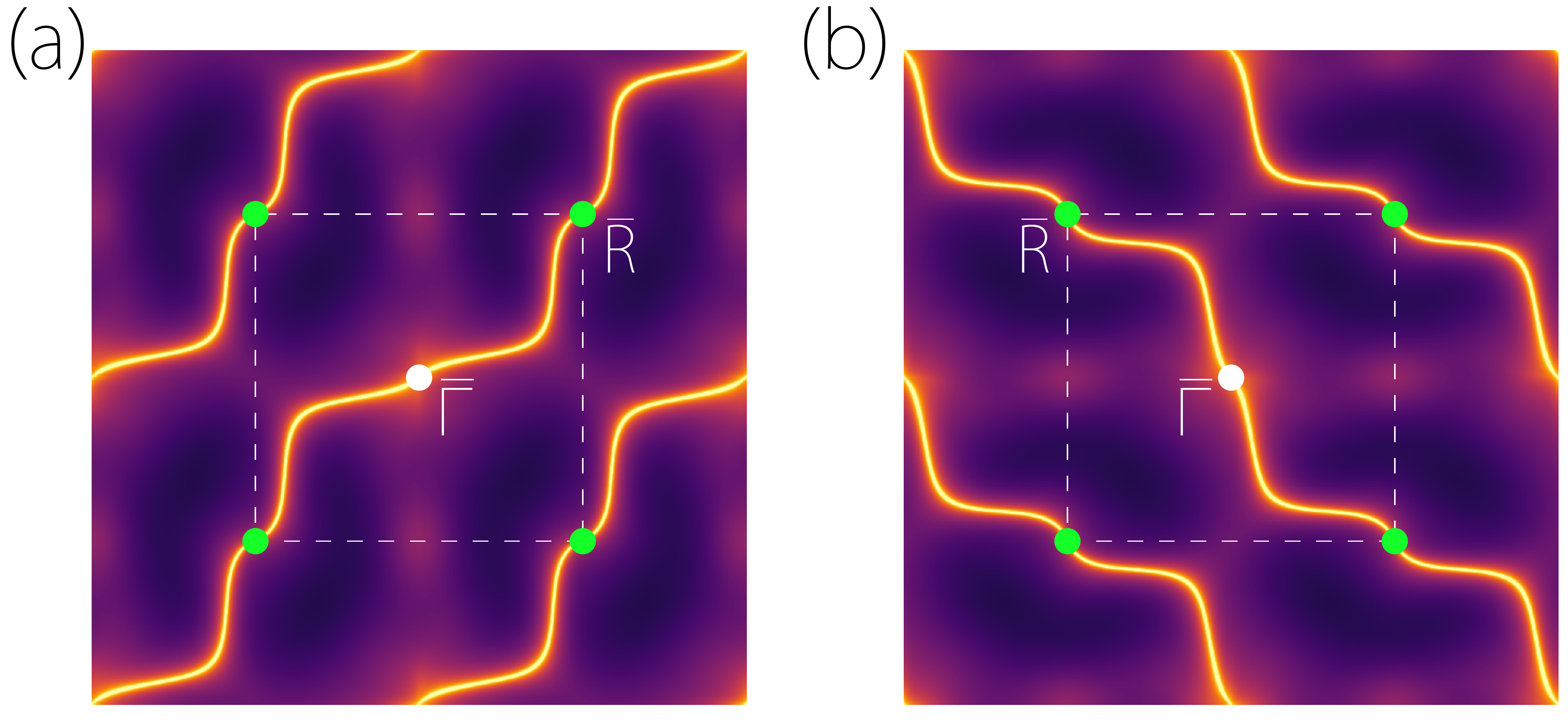}
	\caption{Surface states for the lattice model (\ref{tb195}) with SG 195 on the surface (a) $(001)$ and (b) $(010)$. The white (green) dot denotes the projection of TP at $\Gamma$ (R) point. The model parameters are the same as in Fig.~1.}
	\label{sg195arc}
\end{figure}

According to the bulk-boundary correspondence, a BDP with a nonzero topological charge would generate protected surface states~\cite{Wan2011}. For a Charge-2 TP,  there must be two Fermi arcs emanating from its projection in the surface BZ. To demonstrate this explicitly, we calculate the surface spectra based on the lattice model with SG~195 in Eq.~(\ref{tb195}). From Fig.~\ref{fig:SG195}, we already know that there are two charge-2 TPs with opposite Chern numbers $\pm 2$ at $\Gamma$ and $R$. In Fig.~\ref{sg195arc}(a), we plot the surface spectrum for the (001) surface.
One can clearly find two surface Fermi arcs connecting the projections of the two charge-2 TPs, consistent with the Chern number. Moreover, since the two TPs are pinned at the high-symmetry points $\Gamma$ and $R$, the Fermi arcs must be extensive and traverse the whole surface BZ. Similar observation is also made for the (010) surface, as shown in Fig.~\ref{sg195arc}(b). Such extensive Fermi arcs are desired for experimental study and for possible applications based on Fermi arcs.

\subsection{\label{landaulevel} Chiral Landau Bands}

\begin{figure}[t]
	\centering
	\includegraphics[angle=0, width=0.48\textwidth]{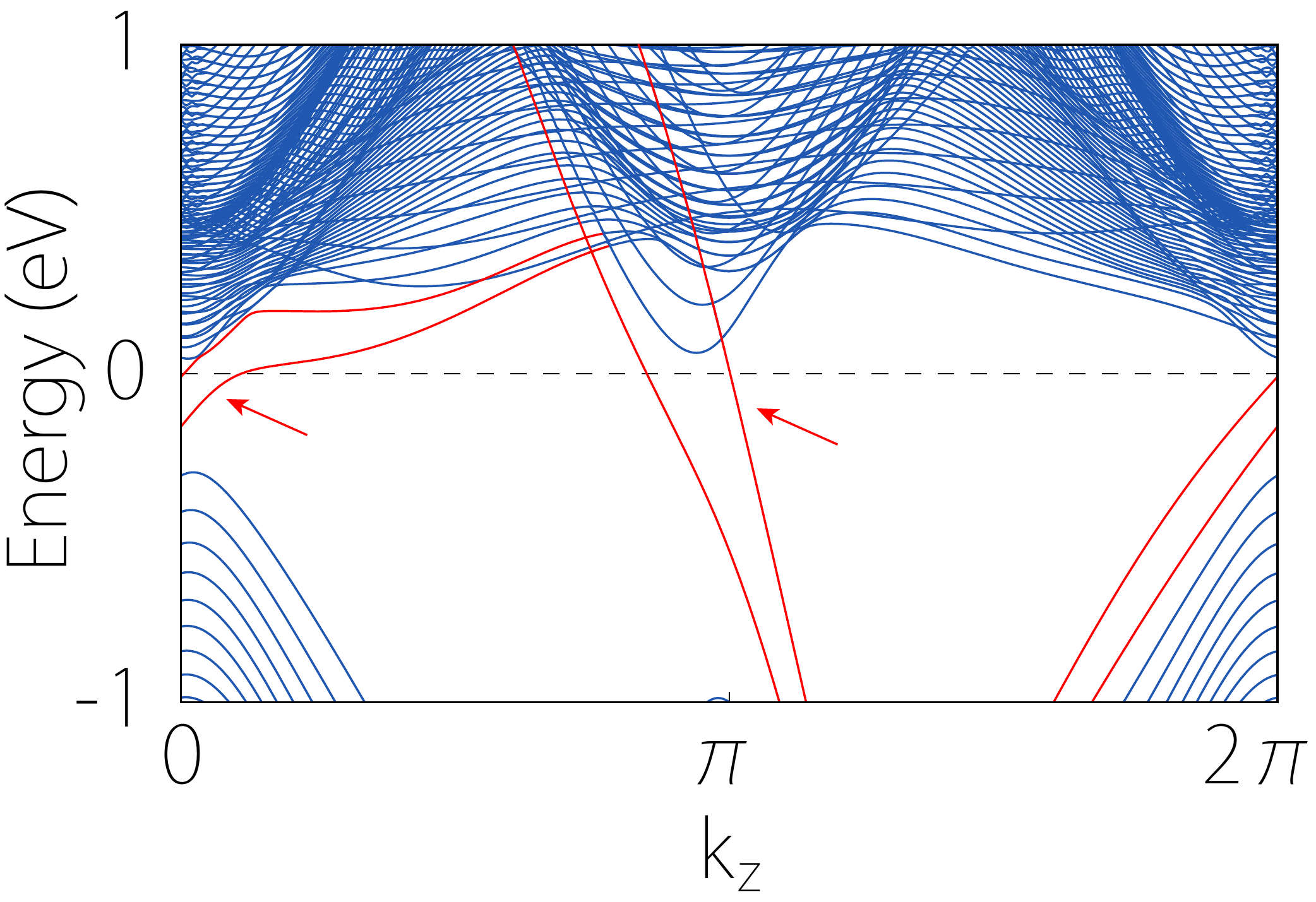}
	\caption{Landau band structure calculated from the lattice model (\ref{tb195}) with SG 195. The magnetic field is along the $z$ direction. The chiral Landau bands are colored in red. }
	\label{landau}
\end{figure}

Under a magnetic field, each Weyl point is featured with a single gapless chiral Landau band, with a definite handedness corresponding to the chiral charge. In a recent work, Zhao and Yang~\cite{Zhao2021} proved a general index theorem, showing that there is an intrinsic connection between the topological charge and the chiral Landau bands for a BDP. It follows that a charge-2 TP should have two chiral Landau bands.
To demonstrate this, we calculate the Landau spectrum for the lattice model in Eq.~(\ref{tb195}) by applying a magnetic field in the $z$ direction. The result is shown in Fig.~\ref{landau}. The electron movement in $x$-$y$ plane is quantized into Landau levels by the $B$ field, so the spectrum consists of 1D Landau bands disperse along the $z$ direction. In the spectrum, one indeed observe two gapless chiral Landau bands around both $k_z=0$ and $k_z=\pi$, which correspond to the locations of the two charge-2 TPs. Such chiral Landau bands may generate intriguing effects, such as chiral anomaly and negative longitudinal magnetoresistance similar to those in Weyl semimetal.

\section{\label{discussion} Discussion and Conclusion}

In Sec.~\ref{chiralTP}, we have discussed the extensive Fermi arcs and the chiral Landau bands for the chiral TPs. In fact, one may expect that many exotic effects proposed for Weyl fermions may also exist for chiral TP fermions, as they both have nontrivial Chern numbers. For example, the unconventional gyrotropic magnetic effect~\cite{Flicker2018}, quantized circular photogalvanic effect~\cite{Juan2017} as well as anomalies in lattice dynamics~\cite{Rinkel2017} should also be investigated in the context of chiral TPs. Particularly, we note that the absence of mirror/inversion/rotoinversion symmetries in chiral SGs will generally make the opposite chiral TPs located at inequivalent energies. This would facilitate the possible observation of the circular photogalvanic effect~\cite{Juan2017}. On the other hand, due to the different pseudospin structure, one can also expect that the chiral TPs could bring new features beyond the Weyl points. This is already evidenced in Ref.~\cite{Shen2010,Fang2016,Feng2020}.

Our results in Table ~\ref{thsp} and \ref{tab:thsl} provide useful guidance for searching or designing concrete systems to achieve the various TPs. For electronic systems, it is promising to search for these TPs in materials made of light elements, such as the carbon and boron allotropes, where SOC can be neglected. Indeed, we have already seen a few examples~\cite{Hu2019,Gao2019}. However, it should also be pointed out that in these examples, the desired TPs are away from the Fermi level and the low energy bands are not clean enough. Hence, it remains an important task to search for ideal candidate materials. Meanwhile, such TPs may also be explored beyond electronic systems. For example, they can be realized in phonon spectra of real materials~\cite{Singh2018}, artificial acoustic/photonic crystals~\cite{Lu2014,Yang2015,Mittal2019,Xue2020}, electric circuit arrays~\cite{Imhof2018,Yu2020}, or even mechanical networks~\cite{Huber2016,Prodan2009}. For artificial systems, we have a huge degree of freedom to tune the various parameters. This will be a big advantage for achieving and studying spinless TPs.

Finally, our analysis here can be extended to systems with broken time reversal symmetry, i.e., for magnetic groups.
Actually, we note that all the four types of TPs found here, including the charge-2 TP, the linear achiral TP, the QCTP and the Quadratic TP, should also exist in systems with broken $\mathcal{T}$, because the $\mathcal{T}$ symmetry itself is not essential in our analysis.

In conclusion, we have systematically investigated all possible TPs in the 230 SGs for spinless systems. We classify all TPs according to their locations in the BZ, their dispersion, and chirality.  Besides the conventional linear achiral TP, we find chiral charge-2 TPs, QCTPs, and quadratic TPs. For each kind of TPs, we present its SGs and low-energy effective models. Lattice models are constructed to explicitly demonstrate the existence of three special TPs. For the charge-2 TPs, we also discuss their physical manifestations in the extensive topological surface Fermi arcs and the chiral Landau bands. Our work provides a comprehensive classification of TPs in spinless systems. It offers useful guidance for exploring TPs in various systems ranging from spinless electronic systems, bosonic systems, to artificial periodic systems.

\begin{acknowledgments}
  The authors thank D. L. Deng for helpful discussions. This work was supported by Singapore Ministry of Education AcRF Tier 2 (MOE2019-T2-1-001), the NSF of China (No. 12004035) and Beijing Institute of Technology Research Fund Program for Young Scholars. We acknowledge computational support from the Texas Advanced Computing Center.
\end{acknowledgments}


\appendix

\section{\label{gell} Gell-Mann Matrices }

Gell-Mann matrices are traceless Hermitian generators of the SU(3) Lie algebra. In this work, the Gell-Mann matrices are taken to be~\cite{GellMann1962}
\begin{align}\nonumber
\Lambda_1=&\begin{bmatrix}
0 & 1 & 0 \\
1 & 0 & 0 \\
0 & 0 & 0
\end{bmatrix},~\Lambda_2=\begin{bmatrix}
0 & -i & 0 \\
i & 0 & 0 \\
0 & 0 & 0
\end{bmatrix},~\Lambda_3=\begin{bmatrix}
1 & 0 & 0 \\
0 & -1 & 0 \\
0 & 0 & 0
\end{bmatrix}, \\\nonumber
\Lambda_4=&\begin{bmatrix}
0 & 0 & 1 \\
0 & 0 & 0 \\
1 & 0 & 0
\end{bmatrix},~\Lambda_5=\begin{bmatrix}
0 & 0 & -i \\
0 & 0 & 0 \\
i & 0 & 0
\end{bmatrix},~\Lambda_6=\begin{bmatrix}
0 & 0 & 0 \\
0 & 0 & 1 \\
0 & 1 & 0
\end{bmatrix},~\\\nonumber
\Lambda_7=&\begin{bmatrix}
0 & 0 & 0 \\
0 & 0 & -i \\
0 & i & 0
\end{bmatrix},~\Lambda_8=\frac{1}{\sqrt{3}}\begin{bmatrix}
1 & 0 & 0 \\
0 & 1 & 0 \\
0 & 0 & -2
\end{bmatrix}.
\end{align}
These matrices, together with the $3\times 3$ identity matrix, form a complete basis for $3\times 3$ Hermitian matrices.

\bibliography{TDP2}

\end{document}